\documentclass[preprint2]{aastex}
\usepackage[greek,english]{babel}

\shorttitle{A proto-cluster at $z=2.45$}
\shortauthors{Diener et al.}

\begin{document}

\title{A proto-cluster at $z=2.45$}
\author{C. Diener\altaffilmark{1,2},
S.J. Lilly\altaffilmark{1}, 
C. Ledoux\altaffilmark{2}, 
G. Zamorani\altaffilmark{3},
M. Bolzonella\altaffilmark{3},
D. N. A. Murphy\altaffilmark{4},
P. Capak\altaffilmark{5},
O. Ilbert\altaffilmark{6},
H. McCracken\altaffilmark{7}
}

\altaffiltext{1}{Institute for Astronomy, Department of Physics, ETH Zurich, Zurich 8093, Switzerland}
\altaffiltext{2}{European Southern Observatory, Alonso de C\'ordova 3107, Casilla 19001, Vitacura, Santiago, Chile}
\altaffiltext{3}{INAF, Osservatorio Astronomico di Bologna, Via Ranzani 1, 40127, Bologna, Italy}
\altaffiltext{4}{Instituto de Astrof\'isica, Facultad de F\'isica, Pontificia Universidad Cat\'olica de Chile, Av. Vicu\~na Mackenna 4860, 782-0436 Macul, Santiago, Chile}
\altaffiltext{5}{Spitzer Science Center, 314-6 Caltech, Pasadena, CA 91125, USA}
\altaffiltext{6}{Aix Marseille Universit\'e, CNRS, LAM (Laboratoire d'Astrophysique de Marseille) UMR 7326, 13388, Marseille, France}
\altaffiltext{7}{Institut d'Astrophysique de Paris, UMR 7095 CNRS, Universit\'e Pierre et Marie Curie, Paris, France}

\begin{abstract}
We present the spectroscopic confirmation of a $z=2.45$ proto-cluster. Its member galaxies lie within a radius of 1.4\,Mpc (physical) on the sky and within $\Delta v \pm 700$\,km/s along the line of sight. We estimate an overdensity of 10, suggesting that the structure has made the turn-around but is not assembled yet.
Comparison to the Millennium simulation suggest that analogous structures evolve into $10^{14}-10^{15}$\,M$_{\odot}$/h type dark matter haloes by $z=0$ qualifying the notion of "proto-cluster". The search for the complete census of mock progenitor galaxies at $z\sim2.5$ of these massive $z=0$ mock clusters reveals that they are widely spread over areas with a diameter of 3-20\,Mpc. This suggest that the optical selection of such proto-clusters can result in a rich diversity regarding their $z=0$ descendants. We also searched for signs of environmental differentiation in this proto-cluster. Whilst we see a weak trend for more massive and more quiescent galaxies within the proto-cluster, this is not statistically significant.

\end{abstract}


\section{Introduction}

The identification of galaxy groups and clusters in the high redshift Universe may offer insights into both the formation of structure in the Universe and the evolution of individual galaxies. The study of the most massive structures at a given epoch serves as a laboratory for cosmology. Also it is known that, at least at later epochs $z<1$, the cluster/group environment can influence the member galaxies through a variety of processes. The existence of a morphology-density relation has been established, stating that denser environments host a higher fraction of morphological types that are typically associated with lower star formation rate (Oemler 1974, Dressler 1980, Balogh et al. 2004, Wuyts et al. 2011). Furthermore the fraction of galaxies that are ''quenched", i.e. in which star-formation has ceased or has been suppressed to yield a specific star-formation rate that is below the inverse Hubble time, is higher in high density environments and amongst satellite galaxies relative to central galaxies at the same mass (e.g. Peng et al. 2010, Peng et al. 2012, Knobel et al. 2013, Wetzel et al. 2013, Kovac et al. 2014, Koyama et al. 2014).
A variety of effects such as ram pressure stripping (Gunn \& Gott 1972, Dressler 1980, Abadi et al. 1999), strangulation (Larson et al. 1980, Kawata \& Mulchaey 2008), harassment (Moore et al. 1996) etc. have been invoked as causes of the suppression of star-formation in satellites.

The terminology of the membership of forming structures at high redshift should be carefully defined. Following Diener et al. (2013), when we refer to an association of galaxies as a cluster (or group) we mean that its member galaxies occupy the same dark matter (DM) halo at the time we observe it. This effectively means that the galaxies lie within the $r_{200}$ perimeter of a single DM halo. Of course, this perimeter cannot be observed directly in the sky, and so reliance must be made on comparison with mock catalogues of galaxies that have been generated from large scale numerical simulations like the Millennium simulation (Springel et al. 2005, Kitzbichler et al. 2007, Henriques et al. 2012). In contrast, the member galaxies of a proto-cluster (or proto-group) are occupying different dark matter haloes at the epoch at which they are being observed, but will later accrete into a common halo by $z=0$. The galaxy members of a proto-group are therefore mostly still the dominant galaxies in their individual dark matter haloes (i.e. are ''centrals") but will later become ''satellites" in the larger structure.

In a similar manner to group/cluster identification via mock catalogues, also proto-clusters can be identified in simulations (Diener et al. 2013 and this work). Furthermore simulations can be used to follow the evolution of a proto-cluster and predict its "product" at any later cosmic time. In turn this approach also provides information about the progenitors of todays clusters.

Whereas the aforementioned environmental processes take place and have been observed in assembled groups and clusters at $z<1$, it is still unclear at which stage of the evolution of a proto-cluster to a cluster the onset of environmental differentiation happens.

It is clear that, at a given stellar mass, the properties of satellites in the local Universe are systematically different from those of typical centrals. This central/satellite differentiation has been established out to $z \sim 1$ (Kovac et al. 2014, Knobel et al. 2013). Within the group environment itself centrals and satellites respond in the same manner to environmental influences. The observed differences arise from most centrals being singletons within their DM halo (Knobel et al. 2014). In analogy to this scenario it is possible that at $z \sim 2$ the members of a proto-group or proto-cluster would not be environmentally differentiated from the field population, since these galaxies will (by definition) still be centrals and not satellites. Whether this is true, is however not clear yet, and a $z<1$ relation does not necessarily hold at $z>2$. Also environmental differentiation (even amongst centrals) could enter in new ways at high redshifts. It is clear in the continuity approach of Peng et al. (2010) that quenched galaxies first appear as the characteristic M$*$ of galaxies (and haloes) approaches the mass-scale at which quenching occurs, which appears to be more or less constant with redshift (Peng et al. 2010, Behroozi et al. 2013). As these galaxies quench, the galaxy mass function decouples from the halo mass function, and subsequently evolves "vertically" (with increasing $\phi*$ but more or less constant M$*$). Lilly et al. (2013) referred to this transition as that between Phase 1 and Phase 2 of the evolution of the galaxy population. Seen another way, the relative numbers of quenched and star-forming galaxies around the galaxy stellar M$*$ reflects the slope of the halo mass function at and above this quenching mass (see the discussion in Birrer et al. 2014). At high redshift, the halo mass function has a Schechter M$*$ that is much closer to this quenching mass than it is at low redshifts, where the dark matter M$*$ is of course much larger. Differences in the halo mass function in different large scale environments may then lead to significant environmental differentiation amongst the population of centrals, quite independent of those astrophysical effects on the group/cluster members which appear to dominate at lower redshifts. 

The literature to date shows at times contradictory examples for environmental influences in proto-clusters at $z>2$. Kodama et al. (2007) detect a well populated emerging red sequence in three $z>2$ proto-clusters suggesting the appearance of massive elliptical galaxies whilst Hatch et al. (2011) only see a poorly populated red sequence in their sample of six proto-clusters at $z\sim2.4$. Furthermore Hatch et al. (2011) find evidence that proto-cluster members are both about twice as massive and have lower specific star formation rates than the field galaxies at the same redshift. A similar, tentative, result was found by Lemaux et al. (2014) in a $z=3.3$ proto-cluster. Shimakawa et al. (2014) on the other hand report increased star formation in two $z>2$ proto-clusters. In our previous work (Diener et al. 2013) we studied lower mass structures than the above mentioned and did not find any evidence for environmental differentiation. The same result is also found by Cucciati et al. (2014) at slightly higher redshift ($z=2.9$). Whilst these different results may have its roots in a variety of causes (f.e. different halo masses) it is also possible that the cause is the proto-cluster selection made by these authors. 

In this work we present a $z=2.45$ proto-cluster (Section 2) that we have identified in a follow-up of a number of proto-group structures originally identified in the zCOSMOS-deep survey (Lilly et al. in prep.) by Diener et al. (2013).
The layout of the paper is as follows: we first describe in Section 2 the follow-up spectroscopic observations that led to the confirmation of the $z=2.45$ proto-cluster. We then compare the distribution of the members of this structure with simulations in Section 3, in order to establish at which stage of the process of cluster assembly it is and to predict its evolution to $z=0$.
In Section 4, we then examine its galaxy population in the search of any differences to the field population at the same redshift. We summarize our results and draw conclusions in Section 5.

All magnitudes are quoted in the AB system and we use the $\Lambda$CDM cosmology with $\Omega_{m} = 0.25, \Omega_{\Lambda}=0.75$ and H$_{0}=73$\,km\,s$^{-1}$\,Mpc$^{-1}$ in line with the parameters used for the Millennium simulation.
We refer to physical (comoving) distances with a trailing "p" ("c"), i.e. pMpc would correspond to physical Mpc.

\section{Data}
\subsection{zCOSMOS-deep and the proto-group catalogue}
The zCOSMOS-deep sample (Lilly et al. 2007, Lilly et al. in preparation) provides around $\sim3500$ spectroscopic redshifts at $2<z<3$, observed with the VLT/VIMOS low resolution LR-Blue grism. This instrument configuration yields a spectral resolution of $\rm{R}=180$ and a spectral range of $3700-6700$\,\AA. The zCOSMOS-deep targets lie in the central field of COSMOS, covering in total a region of $0.92^{\circ}\times0.91^{\circ}$, centered on 10 00 43 (RA), 02 10 23 (Dec), with a denser sampled inner area of $0.60^{\circ}\times0.62^{\circ}$. The targets were selected with a combination of BzK and ugr selection (see Daddi et al. 2004, Steidel et al. 2004). All targets have $B_{AB}<25.25$ and the BzK selected galaxies also fulfill $K_{AB}<23.5$.
The zCOSMOS-deep survey has a sampling rate of $\sim50\%$, consisting of a spatial sampling rate of $67\%$ and a success rate in assigning redshift of $60\%$ (see Lilly et al. in prep. for details). 

In our previous paper (Diener et al. 2013) we identified 42 candidate proto-groups at $1.8<z<3$ in the zCOSMOS-deep sample, using a friends-of-friends (FOF) approach with linking lengths $\rm{dr}=500$\,kpc and $\rm{dv}=700$\,km/s. These proto-groups have $3-5$ members and, as argued in that paper, are not likely to be already assembled at the epoch of observation, but the vast majority of them will assemble by $z=0$.
 We selected seven of these spectroscopically identified proto-groups for follow-up spectroscopy to
confirm the previous member galaxies and to identify additional members.

\subsection{FORS2 data}
The VLT/FORS2 spectroscopy was taken in March 2011, using the instrument in its MXU mode with the E2V CCD being sensitive in the blue ($<5000$\,\AA) and the 300V grism. We observed in total of 5 multi-slit masks and obtained spectra for 114 $z_{phot}\gtrsim 2$ galaxies in five $6.8'\times6.8'$ regions (at times overlapping). The proto-cluster we present in this work was covered by three of the five masks. Two of these were observed for 5.5\,h and the remaining for 6h under good seeing conditions ($0.8'' - 1.0''$).

The targets for the observations were selected from the COSMOS photo-z sample (Capak et al. 2007, Ilbert et al. 2009) as follows:
\begin{enumerate}
\item They had to lie in the surroundings (within 2\,Mpc physical)
of the already spectroscopically confirmed proto-groups.
\item Their photo-z had to be consistent with the respective proto-group redshift (i.e. with a $\Delta v<20'000$\,km/s).
\item The targets had to fulfill $B_{AB}<25.5$ or IRAC 3.6$\mu\rm{m}<22$ (or both).
\end{enumerate}
These new targets were supplemented by the already spectroscopically
confirmed members from zCOSMOS-deep, in order to confirm their redshifts and obtain more accurate relative velocities with the higher resolution of FORS2 in comparison to VIMOS.

The data were reduced in the standard way with the IRAF apextract
package and the redshifts were determined through a visual inspection of the individual spectra. Of the 114 targets, we were able to assign spectroscopic redshifts to 67 objects (or 60\%). The success rate in assigning redshifts was dependent on observing conditions and integration time. As the masks covering the area of interest in this paper had both the best conditions and highest integration times (5.5h and 6h) the actual success-rate in that area is as high as 71\%.

\subsection{Proto-cluster at $z=2.45$}
We detected a large structure with a total of eleven\footnote{With 7 targeted proto-groups and 114 observed objects, we observed $\sim16$ candidate members per proto-group. All proto-groups were confirmed, some with 1-2 additional members. The proto-cluster presented in this work is by far the most extreme structure we found. The overall low rate of additional members is due to the high photo-z uncertainties.} spectroscopically confirmed members at a mean redshift of
$z=2.45$, $\rm{RA}=150.00$ and $\rm{Dec}=2.24$ in the FORS2 data. A list of the members is given in Table 1. The eleven members of this structure all lie within a 1.4\,Mpc radius (physical) on the sky and within a velocity range $\Delta v$ of $\pm 700$\,km/s.

We calculated the root-mean-square (r.m.s.) radial size $r_{rms}$ and velocity spread
$v_{rms}$ to be
$r_{rms} = \sqrt{\sum_{i} r_{i}^2/(N-1)} = 902$\,kpc and
$v_{rms} = \sqrt{\sum_{i} v_{i}^2/(N-1)} = 426$\,km/s, where
$r_{i}$ and $v_{i}$ indicate the distance and the velocity of a galaxy relative to the mean, and N is the number of galaxies. As we will argue in paragraph 3.1 and 4.1, this structure is probably not yet gravitationally bound and so these values should not be used to infer a virial mass of the structure.

\begin{figure}[h!]
\includegraphics[scale=0.38]{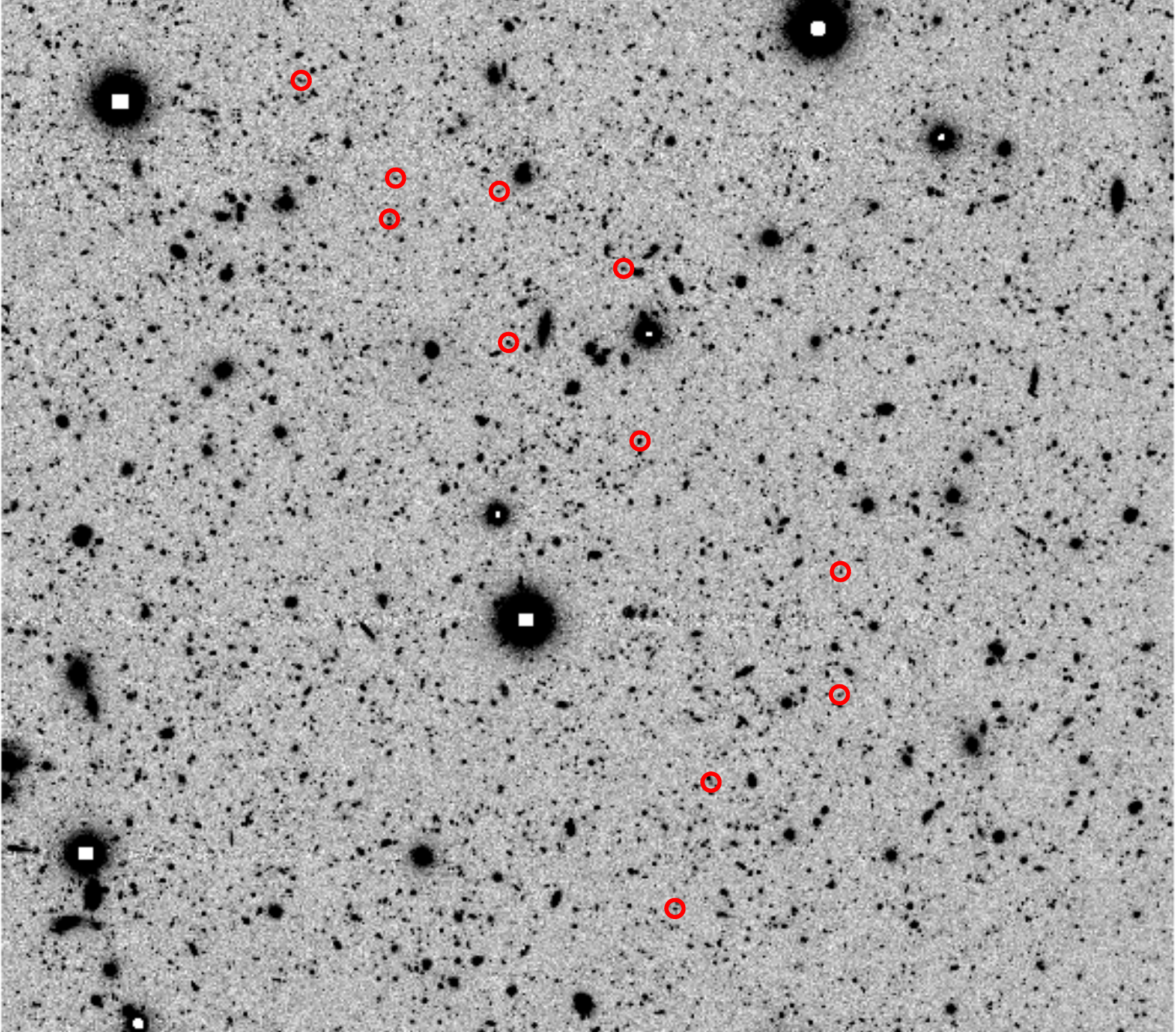}
\caption{The 11 members of the proto-cluster (red circles) in a Subaru B band image. They lie within a radius of 1.4\,Mpc (physical).}
\label{fig_map}
\end{figure}

The spatial distribution of member galaxies is shown in Fig. \ref{fig_map}.

\subsection{The mock sample}
In order to learn about the likely nature of the underlying dark
matter structure of this proto-cluster we need mock catalogues generated to resemble as accurately as possible the observational situation. For this purpose we make use of the Millennium simulation (Springel et al. 2005) and light cones from Henriques et al. (2012). Through the identification of similar structures in these mock samples, where full dark matter and evolutionary information is available, we get indications about the nature and evolution of the observed structure.

\subsubsection{The Millennium simulation \& Henriques lightcones}
The Millennium simulation was following the dark matter distribution in a cubic box of 500\,Mpc\,h$^{-1}$ sidelength starting at $z=127$ and following the evolution of the dark matter particles through time down to $z=0$. As the results are stored in 63 snapshots only, the dark matter structure and its merger trees are built up in a post-processing (Springel et al. 2005, Lemson et al. 2006).
The identified haloes are populated with galaxies whose evolution is described by a semi analytic model (SAM). 

Henriques et al. (2012) construct their light cones from the Millennium simulation volume with the implementation of the SAM described in Guo et al. (2012). They follow the description by Kitzbichler \& White (2007) for periodic replication of the simulation box needed to achieve cones that cover a wide redshift range and assign galaxy redshifts according to the comoving distance of galaxies to a $z=0$ virtual observer.
The resulting 24 lightcones cover an area of 1.4$\times$1.4\,deg$^2$ each.

\subsubsection{Construction of mock samples}
The targets for the FORS2 observations were selected from a photo-z
sample, but chosen to be at the position of known overdensities from the initial spectroscopic zCOSMOS sample. In attempting to mimic this situation as accurately as possible we chose a two-stage approach in constructing the mock sample.

First we created mock samples that were intended to replicate the zCOSMOS-deep sample from which we draw our original candidate group. In this we followed the prescription in Diener et al. (2013), using magnitude cuts on the mock galaxies to achieve number densities in the mocks that match those in the spectroscopic sample. The roughly 50\% sampling of zCOSMOS-deep allows us to construct two mock catalogues from each light cone, resulting in 48 zCOSMOS-deep mock catalogues. Since the proto-cluster in question was originally identified with five zCOSMOS-deep galaxies, we next constructed a mock group catalogue from these zCOSMOS mock samples by applying the same group-finding criteria as for the original candidate group, i.e. we applied a FOF-algorithm with linking lengths $\rm{dr}=500$\,kpc and $\rm{dv}=700$\,km/s, and restricted ourselves to proto-groups with five members.

In a second stage we aimed at reproducing the subsequent FORS2
observations by first creating a mock target sample from the light cones that resembles the underlying COSMOS photo-z sample from which the targets were selected. As mentioned above, the objects in our target catalogue had to fulfill $B_{AB}<25.5$ and/or IRAC $3.6\mu$m$<22$, as well as having a photo-z consistent with the respective previously identified group.
We applied a photo-z error of 10'000km/s (this corresponds to the typical observed photo-z error at $z\sim 2.5$) to the mock redshifts and cut the mock sample in B and IRAC $3.6\mu$m. These cuts where adjusted such that the number density of our target catalogue from COSMOS matched the mock sample. From this mock target catalogue we then randomly draw 16\% of all objects to mimic the product of the fraction of targets actually observed (22\%) and the success rate in assigning redshift (71\%). We populated the already existing group-catalogue with this "observed" sample.

In the final sample we searched for proto-clusters that had 11 or more members lying within 1.4\,Mpc and 700\,km/s (same as the FORS2 proto-cluster). 
This resulted in 16 candidate proto-clusters in the redshift range $2.3<z<2.6$, distributed over the 48 mock samples of $2\deg^2$ each. 

\section{Evolution in simulations}
\subsection{Surface number densities}
As mentioned above, with our selection technique, we detect 16 candidate proto-clusters in the 96\,deg$^2$ of the 48 mock samples, or 0.17 proto-clusters per deg$^2$. In other words, we expect one such system at $z\sim2.5$ redshift range in a 6\,deg$^2$ field. Based on this, to find one in the region of zCOSMOS-deep (1\,deg$^2$ in total and 0.36\,deg$^2$ in the area of maximum coverage) appears lucky, but not exceptionally so.

\subsection{Assembly history}
Whilst at low redshift galaxy clusters will usually have mostly
assembled (i.e. have their member galaxies occupying the same DM halo) and will in many cases be virialised, this is not the case at $z>2$. The growth of structure is so rapid at these masses at high redshift that even quite substantial overdensities will most likely be at a "pre-assembly" stage, meaning that their member galaxies will accrete onto a common dark matter halo by $z=0$ but are still occupying different haloes when they are being observed (e. g. Diener et al. 2013). We refer to these forming structures as "proto-groups" or "proto-clusters".

We can use the properties of the structures in the mock catalogues
to infer the likely state of the system we see in the sky. In the case of the 16 proto-clusters in the mock sample, the majority (10, or 62.5\%) have already started assembly at $z\sim2.5$, in the sense that the largest halo already contains between two and four galaxies that meet our selection criteria (note that there may also be fainter galaxies residing in the same haloes). About a third of the $z\sim 2.5$ proto-clusters however still consist entirely of singletons. The assembly process continues to $z=0$ when 13 (81\%) have fully assembled (i.e. with all the detected members within a common halo) or mostly assembled (i.e. more than $50\%$ of its members in a common halo). Only for three (19\%) of the mock clusters the contamination by interlopers is high enough that less than $50\%$ of the identified members end up occupying the same halo at $z=0$.

\begin{figure}[h!]
\includegraphics[scale=0.43]{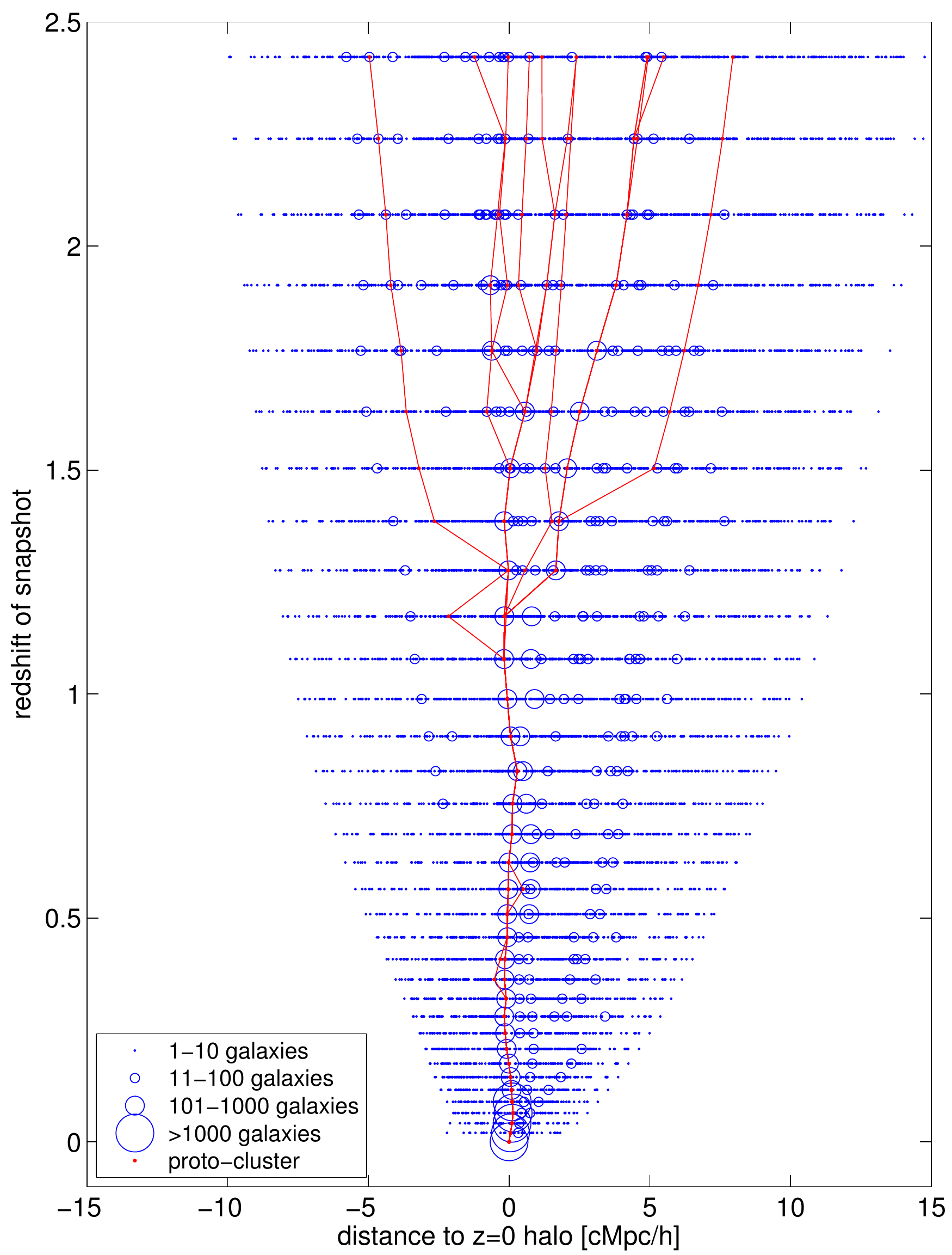}
\caption{We show the assembly history of a $z\sim2.5$ proto-cluster by following all haloes that will eventually become part of the same $z=0$ DM halo, i.e. form a cluster. The size of the circles corresponds to the number of galaxies that inhabit a given halo. Whilst at $z\sim2.5$ galaxies are mostly centrals themselves, they continuously accrete onto other haloes to eventually become satellites in the $z=0$ cluster. The proto-cluster member haloes we identify at $z\sim2.5$ are highlighted in red.}
\label{fig_assembly}
\end{figure}

We illustrate the assembly of such a proto-cluster in Fig \ref{fig_assembly}, by following the haloes of all galaxies from $z\sim2.5$ that will eventually become members of the same $z=0$ cluster. We highlight the proto-cluster galaxies that we identified in our mock-catalogue in red, but obviously many more galaxies are part of this massive $z=0$ cluster and at $z\sim2.5$ they are distributed over rather large scales (see section 3.4 for further discussion). Also evident from this figure is that the originally identified proto-cluster members largely complete their accretion process before $z=1$, consistent with the idea that the structure has made its turn-around (see section 4.2).

Overall, on average, in the mock catalogues, 78\% of the identified
proto-cluster members will end up being true cluster members by the
present epoch whilst only 16\% are already in the same halo at $z\sim 2.5$. These numbers suggest that the presented structure is a true
proto-cluster in the sense that the vast majority of the galaxies will end up in a massive (see next section) cluster today, but only a small minority are already sharing the same dark matter halo at the high redshift that we observe them.

\subsection{Halo masses}
\begin{figure}[h!]
\includegraphics[scale=0.5]{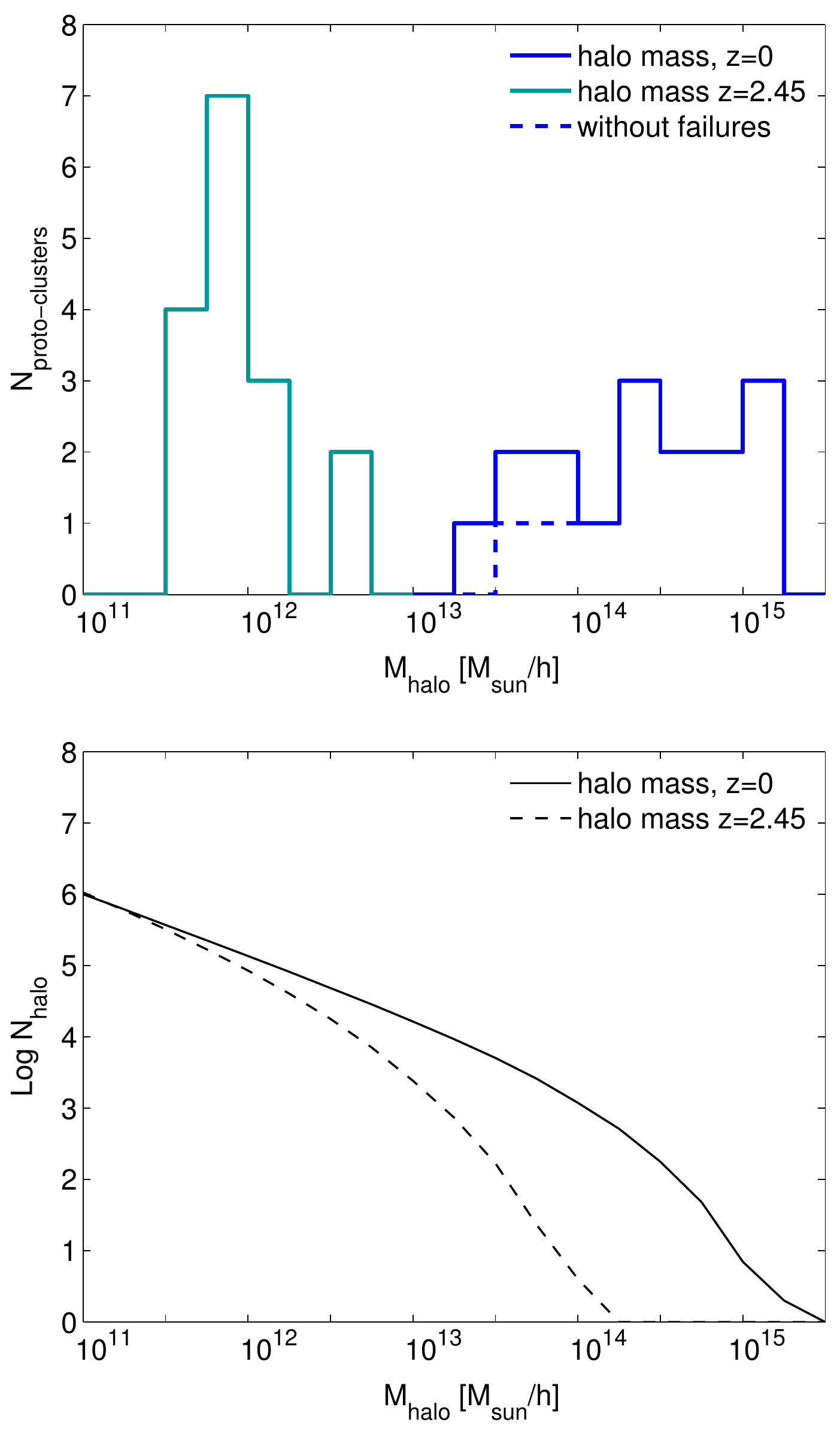}
\caption{Top panel: We show the halo masses of the most massive halo of the mock proto-clusters at $z=2.45$ (turquoise) and at $z=0$ (blue). Whilst evolving from a rather unremarkable halo ($\sim 10^{12}$\,M$_{\odot}/$h) they will become some the most massive clusters by $z=0$ with a halo mass of $\sim 5 \times 10^{14}$\,M$_{\odot}/$h. The dashed line indicates the halo masses without the 3 clusters that do not assemble, i.e. that end with $<50$\% of the members in the same halo.
Bottom panel: We show the halo mass functions at $z=2.45$ (dotted) and $z=0$ (solid) for comparison.}
\label{fig_halomasses}
\end{figure}
As established in the previous section, the member galaxies of the
proto-cluster are not likely to be occupying the same dark matter halo at $z\sim2.5$. However it is illustrative to compare the typical halo at $z\sim2.5$ to the fully evolved cluster halo at $z=0$ by following the evolution of these haloes in the simulation. At $z\sim2.5$ the proto-cluster galaxies are residing in somewhat unremarkable halo masses of $\sim10^{12}$\,M$_{\odot}$/h, simply because they are mostly singleton galaxies. This changes dramatically by $z=0$ when the former proto-cluster members mostly inhabit haloes with M$_{halo} = 10^{14}-10^{15}\,$M$_{\odot}$/h, i.e. they become members of the most massive clusters seen today. This is illustrated in Fig. \ref{fig_halomasses} where we both show the distribution of halo masses of the proto-cluster at $z\sim2.5$ and $z=0$ (top panel) as well as the halo mass function at both redshifts for comparison (bottom panel). 
This again underlines the use of terminology "proto-cluster". At $z\sim2.5$ this structure is an assembly of a galaxies residing as centrals in their DM halo. As it evolves these haloes merge to eventually form a massive cluster halo that is occupied by the previously identified centrals as well as galaxies that accreted later on or were below the detection limit at $z\sim2.5$.

\subsection{Progenitor galaxies}
We have established in the previous section that the mock proto-clusters evolve into very massive $z=0$ clusters. This suggests that other progenitor galaxies to these clusters exist, than the $\sim11$ identified members. All of these progenitors will become part of the same DM halo by $z=0$. They could have failed to be identified as members of the proto-cluster for a variety of reasons. First of all spectroscopy was restricted to relatively bright ($B_{AB}<25.5$) targets. The objects that met the selection criterion were have been sampled incompletely, both due to a limited spatial sampling\footnote{The FORS2 observations only allowed $\sim 20-25$ objects per mask.} and a $<100$\% success rate in assigning redshifts. 

\begin{figure}[h!]
\includegraphics[scale=0.40]{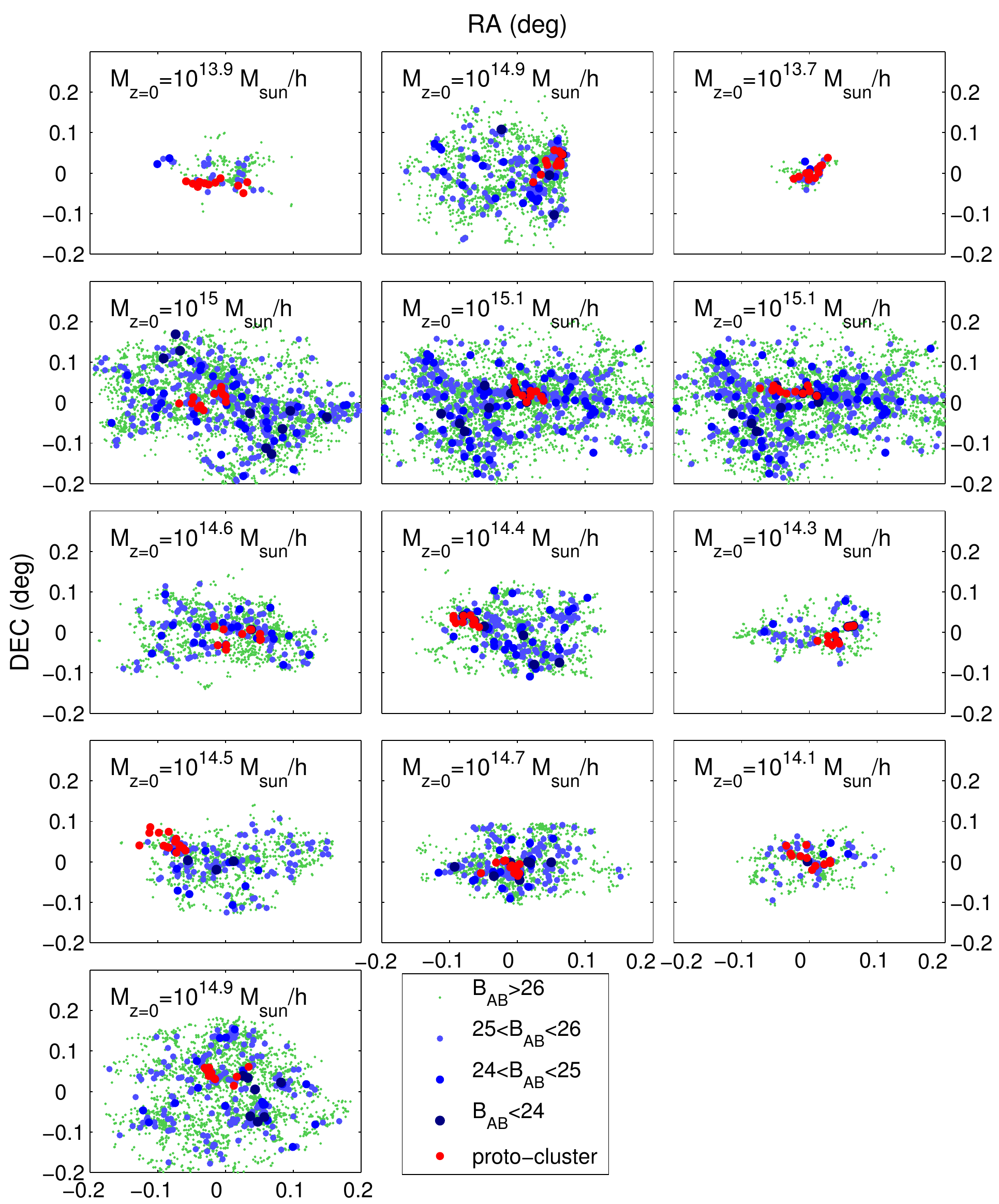}
\caption{We show all $z\sim2.5$ progenitor galaxies (green and blue) that will by $z=0$ become members of the cluster that we identified by our proto-cluster selection. The actual proto-cluster members that identify the structure are highlighted in red. In each proto-cluster field there exist several hundred to thousand progenitors, most of them being too faint for observations. We also note the $z=0$ halo mass which reflects the number of progenitors. Two of the $z=0$ clusters are identical. Reason for that being that their progenitor hosts so many galaxies that they were detected in both of the depleted mock catalogues (we randomly split the original catalogue into two parts to mimic the spectroscopic sampling rate of zCOSMOS-deep.)}
\label{fig_progenitors}
\end{figure}

  We searched for the additional $z\sim 2.5$ progenitors in the lightcones, the result being shown in Fig. \ref{fig_progenitors}. The progenitors are colour- and size-coded according to their B band magnitude, showing the very faint objects in green and the brightest in dark blue.
There is a significant number of such progenitors present in each of the proto-cluster fields (median of 2215, the $z=0$ cluster will have less members than that as some progenitor galaxies merge), however most of them are too faint to have met our selection criterion. The vast majority (95\%) of these objects however meets the $\Delta v < 700$\,km/s condition that would associate them with the proto-cluster if observed.

The diameter of the area occupied by progenitors is ranging from 3\,pMpc to 20\,pMpc. 
 This range of areas is also reflected in the range of halo masses (Fig \ref{fig_halomasses}), which occupy almost 2 orders of magnitudes. Only as the cluster assembles it turns into the more compact structure that is observed at lower redshifts. The optical selection of such a proto-cluster can hence result in a diversity of objects.
This analysis also hints towards a more extended structure at $z=2.45$ in the COSMOS field. As $3\,\rm{pMpc}-20\,\rm{pMpc}$ correspond to an angular scale of $6'-41'$, comparable or bigger than the FORS2 FOV ($6.8'\times6.8'$), we would not have detected this extended structure with our observation.

\section{Observational Characteristics}
According to simulations the $z=2.45$ proto-cluster is likely to evolve into a massive cluster by $z=0$, but is only just starting its assembly. Whilst this implies that the effects which shape the group population at $z<1$ can not take place yet, the overdense environment of a proto-cluster could influence the member galaxies and hence make them distinguishable from the field galaxies at the same redshift.

\subsection{Photo-z samples}
The selection of galaxies in zCOSMOS-deep and also for the FORS2 observation involved a cut in the B magnitude. The spectroscopic sample is therefore highly incomplete in mass and will be biased against red objects that are quiescent or which have high reddening. Any interesting statements about the population of galaxies in the proto-cluster region compared with those in the surrounding ``field'' must therefore be based on photo-z sample(s), despite the high redshift uncertainties therein.

To that end we use two photo-z samples, one being the i-band selected photo-z catalogue from Capak et al. (2007) and Ilbert et al. (2009), to which we applied the same selection criteria as for the FORS2 observations. As this is then exactly the parent sample for the observations, it replicates our selection function. We base an estimate of the overdensity on this sample.

To better address the issue of incompleteness we employ the UVISTA catalogue from McCracken et al. (2012), containing in total 1747 objects in the zCOSMOS-deep field and with $z_{phot}$ (Ilbert et al. 2013) consistent with the proto-cluster redshift. This sample is K-selected with and complete (to 95\%) down to $K_{AB}=23.8$ corresponding to an approximate mass completeness limit of $\sim10^{10}$\,M$_{\odot}$.
We include this sample to look for differences in the galaxy population at the proto-cluster position with respect to the field. As we are only interested in differential effects it is acceptable if our sample is not complete towards lower masses as long as it includes the objects we are interested in. It should however be noted that the UVISTA sample does not necessarily include the known spectroscopic members (in fact it only contains 6 of the 11 members).

\subsection{Overdensity}
We can roughly estimate the overdensity of the proto-cluster by using the parent photometric sample from which we selected the targets for observation.

To that end we calculate the field density in the $0.6^{\circ}\times0.62^{\circ}$ zCOSMOS-deep (densely sampled) area within the redshift range $z_{cl}\pm 0.12$ which corresponds to $\pm$10000\,km/s, to encompass the photo-z uncertainty. Then: $\rho_{field} = \frac{N_{field}}{1/3 \times area \times (l_{max}^3 - l_{min}^3)}$, where l denotes the comoving distance along the line of sight, $l_{min}$ and $l_{max}$ correspond to the distance at $z_{cl}-0.12$ and $z_{cl}+0.12$. The "area" is the area of zCOSMOS-deep ($1.13\times10^{-4}$\,sr$^2$).

When computing the density of the proto-cluster, we correct for the effect of the redshift uncertainty. The $\Delta v \pm 700$\,km/s of the spectroscopic members is presumably overestimating the extent of the proto-cluster along the line of sight. We therefore assume that in reality the excess of objects concentrates within the $r_{phys} = 1.4$\,Mpc radius, both along the line of sight and radially. Hence the density of the proto-cluster is as follows: $\rho_{cl} = \frac{11}{\pi\,r_{com}^2\,l_{com}}$, with $r_{com} = r_{phys}*(1+z_{cl})$ and $l_{com}=2*r_{com}$.
Then the overdensity is given by $\delta = (\rho_{cl}-\rho_{field)} / \rho_{field} = 10$.

We double-checked our assumption of the spectroscopic members being concentrated within a radius of $1.4\,\rm{pMpc}=4.8\,\rm{cMpc}$. To that end we determined the spread in the cosmological redshifts of the 16 mock proto-clusters (being an measure for the "true" distribution of the proto-cluster member galaxies). The average root-mean-square of these redshifts is 0.006, translating to 7.3\,cMpc which is consistent with the 4.8\,cMpc radius from above, suggesting that our assumption was valid.

An overdensity of 10 implies, in line with the simulations, that whilst the proto-cluster is not likely to be gravitationally bound yet, it has made its turn-around.

\subsection{Radio galaxies}
Whilst this proto-cluster has been selected purely through a FOF approach on a spectroscopic sample, it is well established that radio galaxies are beacons for high-z overdensities (see for example Miley et al 2006, Hatch et al. 2011 and others). We searched the publicly  available FIRST catalogue (White et al. 1997) for sources at the proto-cluster position and found a radio galaxy at ($\rm{RA}=150.0025$, $\rm{Dec}=2.2586$) with a flux of 4.21\,mJy. This position coincides with the proto-cluster with an offset of 0.5\,pMpc from the center.
Castignani et al. (2014) also report a structure at $z=2.39$ at our proto-cluster position identified with a poisson probability method using photometric redshifts looking for overdensities around radio galaxies. They associate their structure with the same radio galaxy and quote a photometric redshift of $z_{phot}=2.2\pm^{0.32}_{0.44}$ for it. Given the uncertainty in photometric redshifts it is possible that our proto-cluster and the structure from Castignani et al. is the same overdensity and associated to the FIRST radio galaxy. Without spectroscopy we can however not make a decisive statement.

\subsection{Does environment matter?}
As discussed in the introduction, previous work finds at times contradictory results regarding environmental differentiation in proto-clusters. The proto-cluster presented in this work has originally been selected from a sample of blue star-forming galaxies as opposed to the predominantly H$\alpha$ selected samples of the aforementioned studies. This opens the door for the search of environmental signatures both identical or different.

To this end we search for any differences in the masses, star-formation rates and the quiescent fraction in the proto-cluster. Due to our blue selection we are however biased towards lower mass and star-forming galaxies. To overcome this limitations we rely on the UVISTA catalogue described in section 4.1. 

We determine the fraction of massive ($\rm{M}>10^{10.5}$\,M$_{\odot}$) galaxies, as well as the fraction of highly star-forming ($\rm{SFR}>50$\,M$_{\odot}$/yr) galaxies within the proto-cluster consistent with proposed scenarios of either overabundance of massive galaxies (Hatch et al. 2011) or elevated star-formation (Shimakawa et al. 2014). At the same time we also search for a difference in the quiescent fraction in comparison to the field, akin to low redshift results.

We make use of the masses and SFRs that are given in the catalogue and which are determined by the mass (SFR) of the best fitting template defined by the median of the likelihood distribution from the photo-z fitting procedure. The selection of quiescent galaxies is also taken from UVISTA, where they employ a criterion based on NUV-R/R-J colours. In total 73 galaxies with $z_{phot}$ consistent with the proto-cluster redshift are flagged as quiescent.

 We calculate the respective fractions of massive, star-forming and quiescent galaxies in the proto-cluster in a cylinder of $r=1.4$\,Mpc radius (physical, the proto-cluster radius) and a length of $\pm 10000$\,km/s (to encompass the photo-z uncertainty). 
To compute the field values we put down cylinders of the same volume at hundred random positions in the zCOSMOS-deep field.

\begin{figure}[h!]
\includegraphics[scale=0.42]{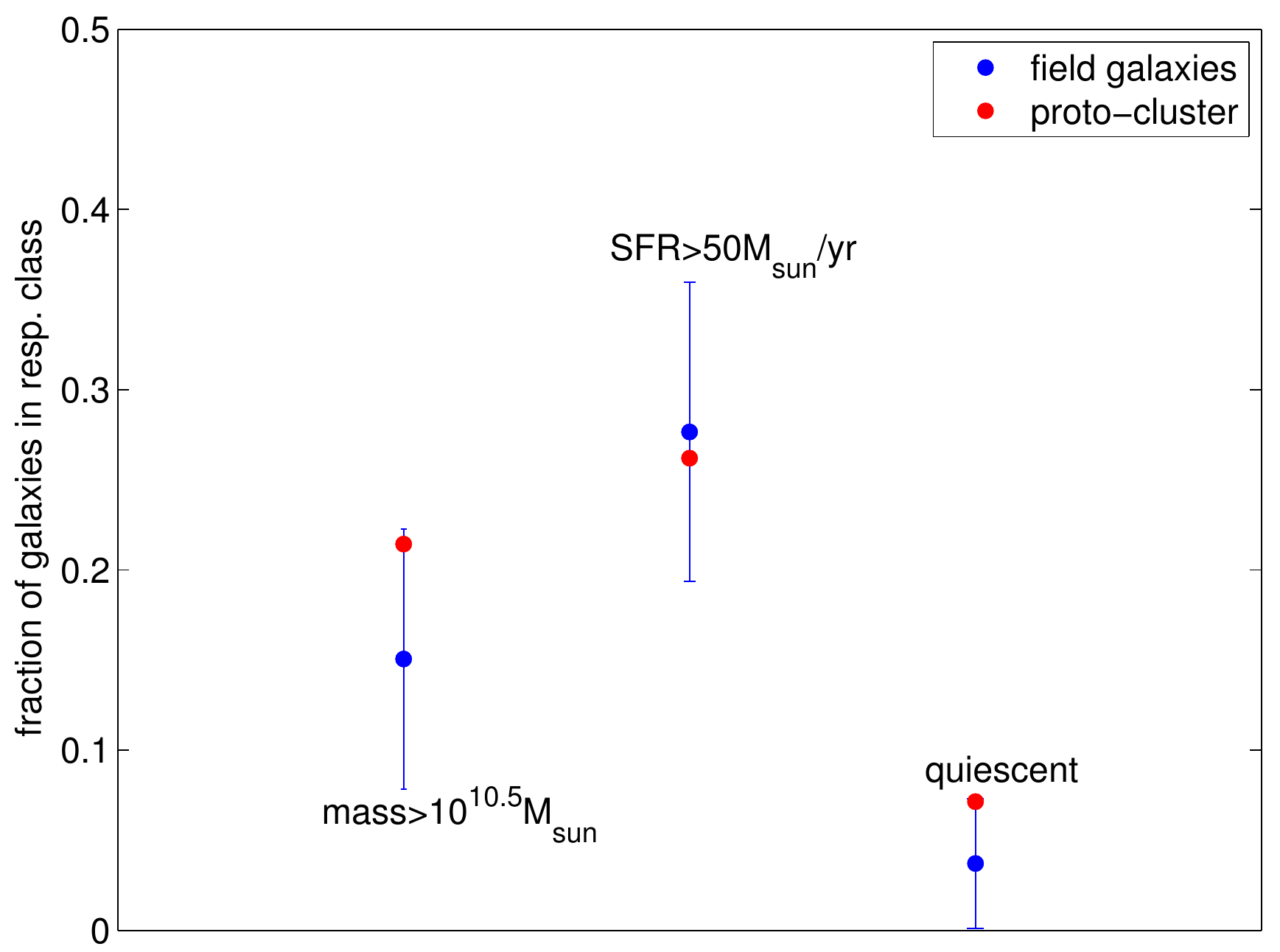}
\caption{We show the fraction of massive $\rm{M}>10^{10.5}$\,M$_{\odot}$ (left), highly starforming $\rm{SFR}>50$\,M$_{\odot}$/yr (middle) and quiescent galaxies in the proto-cluster (red) and the field (blue). There is a weak trend for more massive and quiescent galaxies within the proto-cluster, this is however not significant.}
\label{fig_prop}
\end{figure}

Figure \ref{fig_prop} shows these fractions in comparison with the field: the fraction of massive galaxies left, fraction of star-forming galaxies in the middle and the quiescent fraction right. Whilst we see a trend towards slightly more massive and quiescent galaxies within the proto-cluster, this is not statistically significant within our sample. Despite its likely evolution into a  very massive $z=0$ cluster, we do not see evidence for environmental differentiation at this stage, although it is possible that a weak effect was not detected due to the large errors caused by the use of photo-z.

\section{Summary and conclusions}
We presented a $z=2.45$ proto-cluster with 11 spectroscopically confirmed members. It has first been identified in zCOSMOS-deep and then been followed up with FORS2 spectroscopy. Its member galaxies lie within a radius of 1.4\,Mpc (physical) on the sky and within $\Delta v = \pm 700$\,km/s. We estimated an overdensity of 10, in line with the structure having made the turn-around, but not having accreted its member galaxies onto a common dark matter halo.

This picture is confirmed by comparison of the proto-cluster to similar structures in simulations. To that end we carefully constructed mock catalogues that resemble the observational situation and identified analogous proto-clusters therein.
We follow the evolution of these mock proto-clusters from $z\sim2.5$ to $z=0$. We find that indeed most of the member galaxies are still centrals in their own DM halo at $z\sim2.5$. By $z=0$ most of them share the same halo and hence form a cluster. Furthermore the $z=0$ halo is of $M \gtrsim 10^{14}-10^{15}$\,M$_{\odot}$/h, equivalent to a Virgo or Coma like cluster. 

We identified all $z\sim2.5$ mock progenitor galaxies that will by $z=0$ share the DM halo with the originally identified mock proto-cluster galaxies. These galaxies would mostly be too faint for observations, however still lie within the $\Delta v \pm 700$\,km/s to be associated with the proto-cluster. For each of the mock proto-clusters there exist several hundred to thousand of these progenitors spread over an area with a diameter between 3 and 20\,pMpc and hence are occupying a much wider field than suggested by the originally identified members. This optical selection of a proto-clusters results therefore mostly in lose structures and rich diversity of objects. In order to fully characterise the progenitor population of today's massive clusters these wide fields need to be observed. The numbers from above furthermore hint towards an extended structure in the zCOSMOS field. 

In the last section we studied the galaxy population in the area of the proto-cluster in the search of early signatures of environmental differentiation. We compared the fraction of massive, highly star-forming or quiescent galaxies in the proto-cluster to the field. Whilst we see a weak trend for more massive and quiescent galaxies in the proto-cluster, this is not statistically significant.

\acknowledgments
\section*{Acknowledgements}
This research has been supported by the Swiss National Science Foundation (SNF) and the European Southern Observatory (ESO).
It is based on observations undertaken at the ESO Very Large Telescope (VLT) under the Program 290.A-5160(A). It also uses data products from observations made with ESO Telescopes at the La Silla Paranal Observatory under ESO programme ID 179.A-2005 and data products produced by TERAPIX and the Cambridge Astronomy Survey Unit on behalf of the UltraVISTA consortium.

The Millennium Simulation databases used in this paper and the web application providing online access to them were constructed as part of the activities of the German Astrophysical Virtual Observatory.

\begin{tabular}[t!]{|c|c|c|c|c|c|}
\hline
ID & RA & DEC & z$_{spec}$ & r$_{i}$ [kpc] & v$_{i}$ [km/s] \\
\hline
429950 & 149.997 & 2.257 & 2.442 & 463 & -369 \\
429868 & 150.008 & 2.249 & 2.443 & 321 & -256 \\
410000 & 150.009 & 2.264 & 2.442 & 713 & -322 \\
409614 & 149.995 & 2.24 & 2.439 & 167 & -565 \\
1029209 & 149.976 & 2.227 & 2.44  & 846 & -530 \\
1034036 & 149.992 & 2.194 & 2.451 & 1409 & 414 \\
1031108 & 149.976 & 2.215 & 2.446 & 1064 & -17 \\
1023628 & 150.019 & 2.265 & 2.446 & 891 & 31 \\
1023927 & 150.019 & 2.261 & 2.45	 & 812 & 361 \\
1032336 & 149.988 & 2.207 & 2.453 & 1085 & 655 \\
1022028 & 150.028 & 2.275 & 2.453 & 1278 & 598 \\
\hline 
\end{tabular}

\vspace{5pt}
\textit{Table 1: The 11 spectroscopically confirmed members of the proto-cluster presented in this work. We list their identifier (ID), RA, DEC, redshift as well as radial distance (r$_{i}$) and along the line of sight velocity v$_{i}$ with respect to the proto-cluster center defined by mean RA, DEC and z.}
\twocolumn
\clearpage

\end{document}